\documentclass[prl,10pt,aps,twocolumn,nofootinbib,showpacs,floatfix,amssymb,amsmath]{revtex4-1}
\usepackage{slashed,rotating}

\newcommand{\beqn}{\begin{eqnarray}}
\newcommand{\eeqn}{\end{eqnarray}}
\newcommand{\eq}[1]{(\ref{#1})}

\newcommand{\cL}{{\cal L}}
\newcommand{\cZ}{{\cal Z}}
\newcommand{\cD}{{\cal D}}

\newcommand{\oeps}{\overline{\epsilon}}

\newcommand{\tr}{{\mathrm{tr}}}
\newcommand{\GN}{{\mathrm{GN}}}
\newcommand{\NJL}{{\mathrm{NJL}}}

\renewcommand{\matrix}[4]{\left(\begin{array}{cc} #1 & #2 \\ #3 & #4\end{array}\right)}

\def\bbbone{{\mathchoice {\rm 1\mskip-4mu l} {\rm 1\mskip-4mu l} {\rm 1\mskip-4.5mu l} {\rm 1\mskip-5mu l}}}

\begin{document}

\title{Spontaneous electromagnetic superconductivity of vacuum in strong magnetic field: evidence from the Nambu--Jona-Lasinio model}

\author{M. N. Chernodub}\email{On leave from ITEP, Moscow, Russia.}
\affiliation{CNRS, Laboratoire de Math\'ematiques et Physique Th\'eorique, Universit\'e Fran\c{c}ois-Rabelais Tours,\\ F\'ed\'eration Denis Poisson, Parc de Grandmont, 37200 Tours, France}
\affiliation{Department of Physics and Astronomy, University of Gent, Krijgslaan 281, S9, B-9000 Gent, Belgium}

\begin{abstract}
Using an extended Nambu--Jona-Lasinio model as a low--energy effective model of QCD, we show that the vacuum in a strong external magnetic field (stronger than $10^{16}$ Tesla) experiences a spontaneous phase transition to an electromagnetically superconducting state. The unexpected superconductivity of, basically, empty space is induced by emergence of quark-antiquark vector condensates with quantum numbers of electrically charged rho mesons. The superconducting phase possesses an anisotropic inhomogeneous structure similar to a periodic Abrikosov lattice in a type-II superconductor. The superconducting vacuum is made of a new type of vortices which are topological defects in the charged vector condensates. The superconductivity is realized along the axis of the magnetic field only. We argue that this effect is absent in pure QED.
\end{abstract}



\date{December 28, 2010}

\maketitle

Strong magnetic fields may lead to unusual effects such as magnetic catalysis in the (2+1)--dimensional Gross-Neveu (GN) model~\cite{ref:GN,ref:catalysis:GN}, in QED~\cite{ref:catalysis:QED} and in QCD~\cite{ref:catalysis:QCD}. The strong field supports the chiral magnetic effect in hot quark-gluon plasma~\cite{ref:CME} and a metalliclike conductivity in a quarkless vacuum of lattice SU(2) Yang-Mills theory~\cite{ref:lattice}. Recently, we suggested in Ref.~\cite{ref:I} that an interplay between strong and electromagnetic interactions in a background  of a sufficiently strong magnetic field may turn the cold vacuum into an {\emph{electromagnetic}} superconductor if the strength of the magnetic field exceeds 
\beqn
B_c = m_\rho^2/e \approx 10^{16}\,\mbox{Tesla}\,,
\label{eq:eBc}
\eeqn
where $m_\rho = 775.5\,\mbox{MeV}$ is the mass of the $\rho$ meson and $e$ is the elementary electric charge. Magnetic fields of such strength scale may emerge in the heavy-ion collisions at the Large Hadron Collider in CERN~\cite{ref:Skokov} and, presumably, in the early Universe.

Our idea is based on a very simple argument: the particle spectrum of QCD contains a charged vector resonance (a spin-triplet excitation), $\rho^\pm$ meson, which has a large magnetic dipole moment associated with an anomalous gyromagnetic ratio $g=2$ of the $\rho$ meson. If one treats the $\rho$ meson as a free particle, then in a background of a uniform magnetic field $B$ its ground state energy [corresponding to the Lowest Landau Level (LLL)] becomes a decreasing function of the magnetic field strength, $E_{\rho^\pm}^2(B) = m_{\rho^\pm}^2 - e B$. The energy of $\rho^\pm$ vanishes when the magnetic field reaches the value~\eq{eq:eBc}. As the field strength increases further, the ground state energy $E_{\rho^\pm}$ becomes purely imaginary indicating a tachyonic instability of the ground state towards condensation of the $\rho$ mesons. Since $\rho^\pm$ are electrically charged, their condensation implies an electromagnetic superconductivity of the new ground state. Surprisingly, there is no Meissner effect in the $g=2$ case. Moreover, the strong magnetic field makes the $\rho$ mesons stable at QCD timescales~\cite{ref:I}. 

The suggested vacuum superconductivity has at least two other analogues in particle physics: the Nielsen-Olesen instability of the gluonic vacuum in Yang-Mills theory~\cite{ref:NO}, and the Ambj\o rn--Olesen condensation of the $W$-bosons induced by a strong magnetic field in the standard electroweak model~\cite{ref:AO}. 

In condensed matter physics, a similar phenomenon is known as reentrant superconductivity~\cite{ref:Tesanovic}. Usually, an external magnetic field suppresses the superconductivity via pair breaking effects, so that in a strong magnetic field the superconductivity is lost. However, there are superconductors which may reenter the superconducting phase again at stronger magnetic fields. For example, in the uranium compound URhGe the superconductivity disappears at 2 Tesla, and then, unexpectedly, a stronger  (presumably, spin-triplet) superconductivity reappears again at strikingly large magnetic field of about 8 Tesla~\cite{ref:Uranium}. 

There are various material-dependent proposals to describe specific reentrant superconductors in the condensed matter physics. Our suggestion in QCD~\cite{ref:I} is close to the idea of Refs.~\cite{ref:Tesanovic,ref:reentrant} that in a very strong magnetic field the Abrikosov flux lattice of a type-II superconductor may exhibit a ``reentrant'' quantum regime, characterized by the LLL dominance, the absence of the Meissner effect, a spin-triplet pairing, and a  superconducting current flow along the magnetic field axis only.

In Ref.~\cite{ref:I} we suggested the existence of the new superconducting phase using effective bosonic electrodynamics of the $\rho$ mesons of Ref.~\cite{ref:QED:rho}. However, this model treats the $\rho$ mesons as pointlike particles, and thus it may be inapplicable at strong magnetic fields~\eq{eq:eBc} when the magnetic length becomes of the order of the size of the $\rho$ meson. Here we use the much more general fermionic Nambu--Jona-Lasinio (NJL) model~\cite{ref:NJL} as a low--energy effective theory of QCD in order to show the existence of the electromagnetic superconductivity induced by the strong magnetic fields in the vacuum. 

We consider an extended two-flavor ($N_f = 2$) Nambu-Jona-Lasinio model with three colors ($N_c = 3$) \cite{Ebert:1985kz}:
\beqn
\cL(\psi,\bar\psi)  = \bar \psi \bigl(i \slashed \partial + {\hat Q} \, {\slashed {\cal A}} - \hat M^0 \bigr) \psi + \cL^{(4)}_S + \cL^{(4)}_V\,,
\label{eq:L:NJL}
\eeqn
where the light quarks are represented by the doublet $\psi = (u,d)^T$ and $\hat M^0 = {\mathrm{diag}} (m^0_u, m^0_d)$ is the corresponding bare mass matrix. The uniform magnetic field background $\vec B = (0,0,B)$ is encoded in the Abelian gauge field ${\cal A}^\mu \equiv ({\cal A}^0, {\vec {\cal A}})= (0, - B x_2/2, B x_1/2, 0)$, and the electric charges of the quarks, $q_u = +2e/3$ and $q_d = -e/3$, are combined into 
the matrix $\hat Q = {\mathrm{diag}} (q_u,q_d)$. The hat over a symbol indicates a $2 \times 2$ matrix in the flavor space.

The last two terms in Eq.~\eq{eq:L:NJL} represent the scalar and vector four-quark interactions, respectively:
\beqn
\cL^{(4)}_S & = & \frac{G^{(0)}_S}{2} \bigl[\bigl(\bar \psi \psi\bigr)^2  + \bigl(\bar \psi i\gamma^5 \vec \tau \psi\bigr)^2 \bigl]\,,
\label{eq:L:S}\\
\cL^{(4)}_V & = & - \frac{G^{(0)}_V}{2}  \sum\nolimits_{i=0}^{3} \left[\bigl(\bar \psi \gamma_\mu \tau^i\psi\bigr)^2 + \bigl(\bar \psi  \gamma_\mu \gamma_5  \tau^i \psi\bigr)^2\right]\,,
\label{eq:L:V}
\eeqn
where $G^{(0)}_S$ and $G^{(0)}_V$ are corresponding bare couplings, and $\vec \tau = (\tau^1,\tau^2,\tau^3)$ are the Pauli matrices 

We follow the standard approach~\cite{Ebert:1985kz} and introduce the following bosonic fields corresponding to the quark-antiquark bilinears:  one scalar field $\sigma \sim \bar\psi \psi$, the triplet of three pseudoscalar fields $\vec \pi \sim \bar\psi \gamma^5 \vec \tau \psi$ [made of the electrically neutral, $\pi^0 \equiv \pi^3$, and electrically charged, $\pi^\pm = (\pi^{1}\mp i \pi^{2})/\sqrt{2}$, pions], four vector fields $V_\mu^i \sim \bar\psi \gamma_\mu \tau^i \psi$,
and four pseudovector (axial) fields $A_\mu^i\sim \bar\psi \gamma^5 \gamma_\mu \tau^i \psi$,
\beqn
{\hat V}_\mu & \equiv & \sum\nolimits_{i=0}^3 \tau^i V_\mu^i = \matrix{\omega_\mu + \rho^{0}_\mu}{\sqrt{2} \rho^+_\mu}{\sqrt{2} \rho^-_\mu}{\omega_\mu - \rho^{0}_\mu}\,,
\label{eq:matrix:U}\\
{\hat A}_\mu & \equiv & \sum\nolimits_{i=0}^3 \tau^i A_\mu^i = \matrix{f_\mu + a^{0}_\mu}{\sqrt{2} a^+_\mu}{\sqrt{2} a^-_\mu}{f_\mu - a^{0}_\mu}\,.
\label{eq:matrix:A}
\eeqn

The vector-meson matrix~\eq{eq:matrix:U} is composed of the singlet (in the flavor space) vector (in the coordinate space) $\omega$--meson field $\omega_\mu$, while $\rho^0_\mu \equiv \rho^3_\mu$ and $\rho^\pm_\mu = (\rho^{1}_\mu \mp i \rho^{2}_\mu)/\sqrt{2}$ 
represent, respectively, electrically neutral and charged components of the $\rho$-meson triplet.
The light axial mesons are encoded in the matrix~\eq{eq:matrix:A}: the fields $f_\mu$ and $(a^0_\mu,a^\pm_\mu)$ represent, respectively, the singlet axial $f_1$ meson and the $\vec a_1$ triplet of the axial mesons, respectively.

We rewrite the four-quark interactions~\eq{eq:L:S} and \eq{eq:L:V} via Gaussian integrals over the bosonic fields $\sigma$, $\vec \pi$, ${\hat V}_\mu$, ${\hat A}_\mu$, and integrate over the quarks in the partition function:
\beqn
\cZ {=} \int D \bar \psi  D \psi \ e^{i \int d^4 x \, \cL} {=} \int D \sigma D \pi D V D A \, e^{i S[\sigma,\vec\pi,V,A]}\,,\quad
\nonumber
\eeqn
where the effective bosonic action is
\beqn
& & S[\sigma,\vec\pi,V,A] = S_\psi[\sigma,\vec\pi,V,A]
\label{eq:S}\\
& & \ \ + \int d^4 x \Bigl[- \frac{1}{2 G^{(0)}_S} (\sigma^2 + \vec \pi^2) + \frac{1}{2 G^{(0)}_V} (V^k_\mu V^{k\mu} + A^k_\mu A^{k\mu}) \Bigr] \nonumber\\
& & S_\psi = - i N_c {\mathrm {Tr}}\,  {\mathrm{Ln}}(i \cD)\,,
\label{eq:S:psi} \\
&  & \, i \cD = i {\slashed \partial} + {\hat Q} \, {\slashed {\cal A}} - \hat M^0 + {\hat {\slashed V}}_\mu + \gamma^5 {\hat {\slashed A}} - (\sigma + i \gamma^5 \vec \pi \vec \tau)\,. \quad
\label{eq:icD}
\eeqn

Next, we calculate the effective action~\eq{eq:S} in the strong magnetic field background in the mean field approach. We use simplified notations for the expectation values of the fields, $\langle\sigma\rangle = \sigma$ etc. In the absence of the external magnetic field the expectation values of the fields $\vec\pi$, $V$, and $A$ are zero~\cite{Ebert:1985kz}, while the expectation value of $\sigma$ plays a role of the constituent quark mass, $m_q = \sigma \sim 300$~MeV. 

In order to simplify our calculations, we notice that the presence of the external magnetic field breaks the flavor symmetry down to its diagonal subgroup, so that the diagonal chiral rotations $\Omega = e^{i \alpha_5 \tau^3 \gamma_5}$ can still be used to eliminate the neutral pion condensate $\pi^0$. We also neglect the mass matrix $M^0$ because our calculations show that the conducting properties of the vacuum are almost independent of the {\it {bare}} quark masses $m^0_{u,d} \ll \sigma$. 

The operator \eq{eq:icD} can be represented as the sum $i \cD = i \cD_0 + \hat W$ of the tree-level operator 
$i \cD_0 = i {\slashed \partial} + {\hat Q} \, {\slashed {\cal A}} - \sigma$ and the contribution $\hat W$ from the ``exotic'' condensates,
\beqn
\quad 
\hat W =  {\hat {\slashed V}}_\mu + \gamma^5 {\hat {\slashed A}} - i \gamma^5 (\pi^1 \tau^1 + \pi^2 \tau^2)\,,
\label{eq:hat:W}
\eeqn

At low magnetic fields $\hat W \equiv \langle \hat W \rangle = 0$.
Let us assume that, at certain strong magnetic field $B=B_{\mathrm{c}}^{\NJL}$, the expectation value of the condensate~\eq{eq:hat:W} is nonzero. Let us advance slightly into the new phase taking $B \gtrsim B_{\mathrm{c}}^{\NJL}$, so that the magnitude of the suspected condensate is still small, $ 0 < |\hat W| \ll \sigma$. Then the effective action~\eq{eq:S} can be expanded in powers of the $\hat W$ field, and the fact of the emergence of the new condensate should be seen as a tachyonic instability of the corresponding potential at $\hat W = 0$ (alternatively, we could also solve Schwinger-Dyson gap equations and get the same result).

The tree-level propagator $S^{(0)} \equiv \cD_{0}^{-1}$ of the fermion doublet in the strong magnetic field has the following form:
$S^{(0)}(x,y)  = {\mathrm{diag}} \bigl[S^{(0)}_u(x,y), S^{(0)}_d(x,y) \bigr]$, 
where $S_f$ is the propagator of the $f^{\mathrm{th}}$ quark species. 

The $\rho$-meson condensation and, consequently, the induced superconductivity are the LLL phenomena~\cite{ref:I}. Thus, it is natural to restrict ourselves to the LLL approximation which usually gives a dominant contribution to nonperturbative low-energy quantities in the limit of the strong magnetic field (in particular, because of the large energy gap $\delta E \sim (eB)^{1/2}$ between the LLL and higher levels)~\cite{ref:catalysis:QED,ref:catalysis:QCD,ref:LLL:dominance}. The full solution beyond the LLL approximation will be presented elsewhere. 

In the LLL regime the propagator $S^{(0)}_f$ factorizes into the $B$-transverse and $B$-longitudinal parts which depend, separately, on the $B$-transverse, $x^{\perp} = (x^{1},x^{2})$, and $B$-longitudinal, $x^{\parallel} {=} (x^{0},x^{3})$, coordinates~\cite{ref:catalysis:QED}:
\beqn
S_f^{(0),\mathrm{LLL}}(x,y) = P^\perp_f(x^{\perp},y^{\perp}) \, S^\parallel_f(x^{\parallel} - y^{\parallel})
\label{eq:S:f}
\eeqn
(below we omit the superscripts ``$(0)$'' and ``LLL''). Here
\beqn
P^\perp_f(x^{\perp},y^{\perp}) = \frac{|q_f B|}{2 \pi} 
e^{\frac{i}{2} q_f B \varepsilon_{ab} x^{a} x^{b}
- \frac{1}{4} |q_f B| (x^\perp - y^\perp)^{2}}\,, \qquad
\label{eq:P:perp}
\eeqn
is the transverse projector onto the LLL states and $q_f$ is the electric charge of the $f^{\mathrm{th}}$ quark. 

The longitudinal part of the fermion propagator~\eq{eq:S:f},  $S^\parallel_f \equiv S^\parallel_{{\mathrm{sgn}} (q_f B)}$
is, basically, a fermion pro\-pa\-ga\-tor in the $1+1$ dimensions (we always take $eB>0$ for definiteness), 
\beqn
S^\parallel_f(k_\parallel) = \frac{i}{\gamma^\parallel k_\parallel - m} P^\parallel_f \,,
\qquad
P^\parallel_f = \frac{\bbbone - i f \gamma^1 \gamma^2}{2}\,,
\label{eq:Sf:0}
\eeqn
and the matrix $P^\parallel_f$ (we use $f=\pm1$ for, respectively, $f=u,d$) is the spin projector operator onto the fermion  states with the spin polarized along (for $u$ quarks) or opposite (for $d$ quarks) to the magnetic field. The operator $P^\parallel_f$ projects the original four 3+1 fermionic states onto two (1+1)--dimensional fermionic states, so that fermions can move only along the axis of the magnetic field. The projector~\eq{eq:P:perp} satisfies 
the relation $P^\perp_f \circ P^\perp_f = P^\perp_f$, where  "$\circ$"  is 
the convolution operator in the $B$-transverse space, $A \circ B \equiv {\int} d^2 y^\perp \, A(\dots,y^{\perp}) \, B(y^\perp,\dots)$.

For a coordinate--in\-de\-pen\-dent condensate $\sigma$, the zero-order (in powers of $\hat W$) contribution to the effective action \eq{eq:S:psi} gives us the potential $V(\sigma) = V^{(0)}_\psi(\sigma) + \sigma^2/(2 G^{(0)}_S)$ related to the action as $S = - \int d^4 x \, V$ with  
\beqn
V^{(0)}_\psi = i N_c {\mathrm {Tr}}\,  {\mathrm{Ln}}\, i \cD_0 =
\frac{|eB| N_c}{8 \pi^2} \Bigl[ \sigma^2 \ln \frac{\sigma^2}{\mu^2}  - \Bigl(\frac{1}{\overline{\epsilon}} +1\Bigr) \sigma^2 \Bigr],
\nonumber
\eeqn
where $1/\oeps = 1/\epsilon - \gamma_E + \log 4 \pi$,  $\gamma_E \approx 0.57722$ is Euler's constant, and $\mu$ is a renormalization mass scale. In order to regularize the divergent contributions of the $(1+1)$-dimensional fermions we implemented the dimensional regularization in $d=2 - 2 \epsilon$ dimensions. The renormalization of the NJL coupling constant in the $\overline{\mathrm{MS}}$ scheme, $1/G_S = 1/G^{(0)}_S - N_c |eB|/(4 \pi^2 \oeps)$, resembles the renormalization of the 1+1 dimensional GN model~\cite{ref:GN} 
with the identification $G_S  \equiv 2 \pi G_{\GN}/(N_c |eB|)$ first noticed in~\cite{ref:catalysis:QED}.

The minimum $\sigma = \sigma_{\mathrm{min}}$ of the renormalized potential,
\beqn
V(\sigma)= \frac{1}{2 G_S} \sigma^2  + \frac{|eB| N_c}{8 \pi^2} \Bigl(\ln \frac{\sigma^2}{\mu^2}-1\Bigr)  \sigma^2 \,,
\nonumber
\eeqn
provides us with the $B$-dependent quark mass 
\beqn
m_q(B) = \sigma_{\mathrm{min}}(B) = \mu \exp\{-  2 \pi^2/(G_S N_c |eB|)\}\,.
\label{eq:mq:B}
\eeqn
In the LLL approximation to the NJL model the scale $\mu$ is not fixed as it is related to the $B$-longitudinal $1+1$ motion of the quarks. 
Beyond the LLL approach the scale may perhaps be set as $\mu^2 \propto |eB|$ following Ref.~\cite{ref:catalysis:QCD}.

The effective bosonic model~\cite{ref:I} suggests that the possible superconducting ground state should exhibit an inhomogeneous behavior  in the $B$-transverse plane. Thus, we assume that the exotic condensates may be $x^\perp$-dependent, $\hat W = \hat W(x^\perp)$, and calculate the corresponding quadratic contribution to the effective action~\eq{eq:S:psi},
\beqn
S^{(2)}_\psi = - \int d^4 x\, V^{(2)}_\psi = \frac{i N_c}{2} {\mathrm {Tr}}\,  \frac{1}{i \cD_0} W \frac{1}{i \cD_0} W\,,
\label{eq:V2:psi}
\eeqn
(despite the expected  $x^\perp$--dependence of the condensates $\hat W$, we still call the functional $V^{(2)}_\psi$ as the ``potential''). 

We find that the potential~\eq{eq:V2:psi} involves only the $B$-transverse components of the vector and axial mesons,
\beqn
\int d^2 x^\perp V^{(2)}_{\psi} & = & - \frac{4 N_c |e B|}{ 9 \pi^2} 
\Bigl[\Bigl(\frac{1}{\overline{\epsilon}} - \ln\frac{\sigma^2}{\mu^2} \Bigr)  (\phi^* \circ P_e \circ \phi)  \nonumber \\
& & + \Bigl(\frac{1}{\overline{\epsilon}} - \ln\frac{\sigma^2}{\mu^2} - 2 \Bigr)  (\xi^* \circ P_e \circ  \xi)\Bigr], \quad
\label{eq:V2:bare}
\eeqn
where $\phi = (\rho^+_1 + i \rho^+_2)/2$ and $\xi = (a^+_1 + i a^+_2)/2$.
The $B$-transverse projector for the unit charged particle,
$P^\perp_e(x^{\perp},y^{\perp}) = (9 \pi/|e B|) P^\perp_u(x^{\perp},y^{\perp})  P^\perp_d(y^{\perp},x^{\perp})$
is given by Eq.~\eq{eq:P:perp} with the replacement $q_f \to e$.

The unstable tachyonic mode of the potential \eq{eq:S}, \eq{eq:V2:bare} turns out to be an inhomogeneous eigenstate of the charge-1 projection operator~$P_e$,
\beqn
(P_e \circ \phi)(x^{\perp}) = \phi(x^\perp)\,,
\label{eq:Pe:eigenstate}
\eeqn
The solution is a general Abrikosov-like configuration~\cite{ref:Abrikosov}
\beqn
\phi & = &\phi_0 \, K\bigl({\bar z}/L_B\bigr)\,, \quad L_B = \sqrt{2 \pi/|e B|}\,, \
\label{eq:v} \\
K(z) & = & e^{-\frac{\pi}{2} (|z|^2 + z^2)} \sum\nolimits_{n = -\infty}^{+ \infty} c_n e^{- \pi n^2 + 2 \pi n z}\,,
\label{eq:K}
\eeqn
where $\phi_0$ and $c_n$ are arbitrary complex parameters and $z = x^1 + i x^2$ (and similarly for the axial vector field~$\xi$).

The solution~\eq{eq:v} represents a (periodic) flux-tube structure similar to the Abrikosov lattice which is realized in a mixed state of a type-II superconductor subjected to a near-critical external magnetic field~\cite{ref:Abrikosov}. Generally, the coefficients $c_n$ can be fine-tuned by a complicated minimization procedure if the full potential is known~\cite{ref:Abrikosov}. Here we follow Ref.~\cite{ref:AO,ref:I} and set $c_n=1$ so that the solution~\eq{eq:v} represents a square lattice with the quantized area $2 \pi/|eB| \equiv L_B^2$ given by the magnetic length~$L_B$.

The quadratic potential, evaluated at the solution~\eq{eq:v},
\beqn
V^{(2)} {=} \sqrt{2} \Bigl[\frac{1}{G_B} (|\phi_0|^2 +  |\xi_0|^2) - \frac{2 N_c |eB|}{9 \pi^2} (|\phi_0|^2 -  |\xi_0|^2)\Bigr],
\nonumber
\eeqn
is unstable towards a spontaneous creation of the $B$-transverse $\rho^\pm$ condensates 
with the tachyonic mode $\rho^+_1 = i \rho^+_2 = \phi$ if the strength of the magnetic field exceeds
\beqn
B^{\NJL}_c = \frac{9 \pi^2 }{2 e \,N_c \, G_B}\,, 
\qquad \
\frac{1}{G_B} = \frac{1}{G_V} - \frac{8}{9 G_S}\,,
\label{eq:BNJL}
\eeqn
with $1/G_V = 1/G^{(0)}_V - N_c |eB|/(9 \pi^2 \overline{\epsilon}')$ and $1/\overline{\epsilon}' \equiv 1/\overline{\epsilon} -1$.

Due to the fact that the phenomenological values of the parameters $G_{S,V}$ vary in a broad region~\cite{ref:physical}, and due to subtleties of the renormalization of the effective $(1+1)$-dimensional theory embedded in 3+1 dimensions, we can only give an approximate estimation of the critical field: $e B_c \sim 1\,{\mbox{GeV}}^2$ or $B_c \sim 10^{16} \, {\mathrm {Tesla}}$.

The quartic correction to the potential in Eq.~\eq{eq:S:psi},
\beqn
V^{(4)}_\psi = C_0 \frac{|eB| N_c}{2 \pi^2 m^2} |\phi_0|^4\,, 
\label{eq:V4}
\eeqn
allows us to find the condensate at $B \geqslant B_c^\NJL$:
\beqn
\phi_0(B) = e^{i \theta_0} C_\phi m_q(B) {\bigl(1 - B_c^{\NJL}/B\bigr)}^{1/2}\,,
\label{eq:phi0}
\eeqn
where $\theta_0$ is a constant phase, $C_0 \approx 1.2$, $C_\phi \approx 0.51$ and
the quark mass $m_q$ is given in Eq.~\eq{eq:mq:B}. At $B < B_c^\NJL$ the condensate~\eq{eq:phi0} is zero.
The phase transition at $B=B_c$ is of the second order with the critical exponent~$1/2$.

Thus, the magnetic field induces the quark condensate 
\beqn
\langle \bar u \gamma_1 d\rangle = - i \langle \bar u \gamma_2 d\rangle = \rho_0 (B) \, K\Bigl(\frac{x_1+ i x_2}{L_B}\Bigr) \equiv \rho(x^\perp), \quad
\quad
\label{eq:ud:cond}
\eeqn
where $\rho_0 (B) = \phi_0(B)/G_V$. Using known (see, e.g., Ref. \cite{ref:Abrikosov}) general properties of the function $K(z)$, Eq.~\eq{eq:K}, we conclude that the ground state should be given by a periodic (in general) lattice of a new type of topological vortices which are parallel to the magnetic field. The phase of the condensate \eq{eq:ud:cond} winds around the center of each vortex where the absolute value of $\rho(x^\perp)$ vanishes.

The condensate~\eq{eq:ud:cond} locks the local $U(1)_{\mathrm{e.m.}}$ transformations with the global $O(2)_{\mathrm{rot}}$ rotations of the coordinate space about the magnetic field axis~\cite{ref:I,ref:II}: $U(1)_{\mathrm{e.m.}} \times O(2)_{\mathrm{rot}} \to G_{\mathrm{lat}}$, where $G_{\mathrm{lat}}$ is a discrete symmetry group of rotations of the $\rho$-vortex lattice. 

The new vacuum state is superconducting. One can show that there is no $B$-transverse current, $J^1{=}J^2{=}0$, so that the electric current flows along the magnetic field axis only.  In a very weak (test) electric field $\vec E = (0,0,E_z)$ with $E_z \ll B$, the induced electric current in the new vacuum state~\eq{eq:ud:cond} in a linear-response approximation is (we use the retarded Green functions):
\beqn
J^\mu(x) = \sum\nolimits_{f=u,d} q_f  \langle \bar \psi_f \gamma^\mu \psi_f\rangle \equiv - \tr [\gamma^\mu {\hat Q} S(x,x)]\,, \qquad
\label{eq:Jmu:general}
\eeqn

We average the current~\eq{eq:Jmu:general} over the $B$-transverse plane and,  in the leading order in powers of $\rho$,  we get:
\beqn
\frac{\partial {\mathcal Q}(x^\parallel)}{\partial z} + \frac{\partial {\mathcal J}(x^\parallel)}{\partial t} 
= \frac{2 C_q}{(2 \pi)^3} e^3 \bigl(B - B_c^{\NJL}\bigr) \, E_z \,, \qquad
\label{eq:London:local}
\eeqn
where ${\mathcal Q}$ is the plane-averaged electric charge density $J^0$, ${\mathcal J}$ is the plane-averaged current $J^z$, and $C_q \approx 1$~\cite{footnote}. At $B < B_c$ the right hand side of Eq.~\eq{eq:London:local} is zero. Apart from prefactors, the transport laws in the NJL model~\eq{eq:London:local} and in the $\rho$-meson electrodynamics ~\cite{ref:I} are identical.

The linear-response law~\eq{eq:London:local} can be rewritten in a Lorentz-covariant form, $\partial^{[\mu,} J^{\nu]} =\gamma \cdot  (F, {\widetilde F}) {\widetilde F}^{\mu\nu}$, via the invariants  $(F, {\widetilde F}) =  4 (\vec B, \vec E)$ and $(F, F) = 2 (\vec B^2 - \vec E^2)$. Here ${\widetilde F}_{\mu\nu} = \epsilon_{\mu\nu\alpha\beta} F^{\alpha\beta}/2$ and $\gamma$ is a function of $(F, F)$~\cite{ref:II}.

Equation~\eq{eq:London:local} is a London equation for  an anisotropic superconductivity. Thus, we have just shown that the strong magnetic field induces the new electromagnetically superconducting phase of the vacuum if $B>B_c$. An empty space becomes an anisotropic superconductor.

The superconductivity of the vacuum is a new effect which is realized at the QCD-QED interface. This mechanism should not work in the pure QED since electrically charged spin-1 bound states are absent there. 

\begin{figure}[!thb]
\begin{center}
\includegraphics[scale=0.48,clip=false]{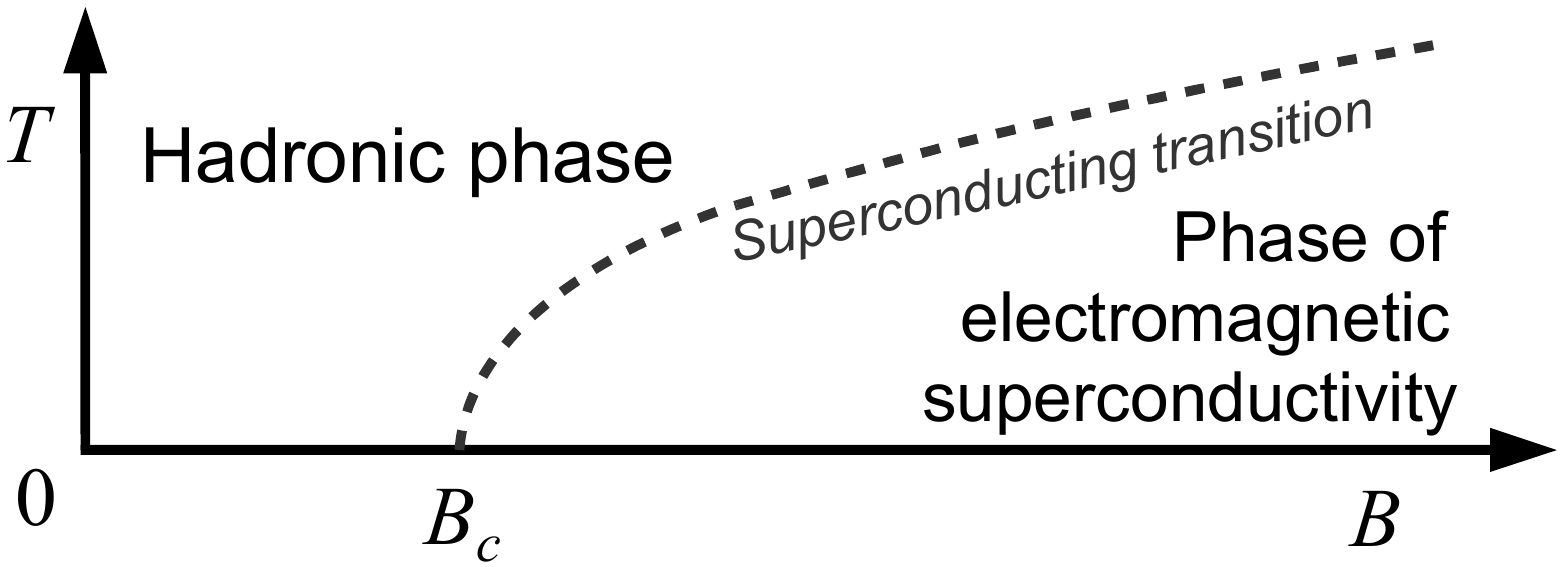}
\end{center}
\caption{Low-temperature part of the QCD phase diagram.}
\label{fig:phase}
\end{figure}

On general grounds one can expect that increase in temperature $T$ (which, in general, should be of a hadronic scale) should lead to an evaporation of the $\rho$ condensate with a loss of the superconductivity. The suggested low-$T$ part of the $B-T$ phase diagram is shown in Fig.~\ref{fig:phase}.

The author is grateful to A.~Nedelin, A.~Niemi, P. Olesen, M.~Ruggieri and  V.~I.~Zakharov for useful discussions. The work was supported by Grant No. ANR-10-JCJC-0408 HYPERMAG.

\end{document}